# Oscillatory superconducting proximity effects in a buried Pt interlayer detected by FMR

L. Peridon[1], T. Courtois[1], S.A Hussain[1], J.B. Tjernshaugen[2], J. Linder[2], S. ElMeligy[3], B. Újfalussy[4], K. Dumesnil[1], S. Mangin[1], S. Petit-Watelot[1], C. Ciccarelli[2,5], T. Hauet[1]

1. Université de Lorraine, CNRS, IJL, F-54000, France
2. Center for Quantum Spintronics, Department of Physics, Norwegian University of Science and Technology, NO-7491 Trondheim, Norway
3. Department of Physics, Virginia Tech, Blacksburg, Virginia 24061, USA
4. Wigner Research Centre for Physics, Institute for Solid State Physics and Optics, H-1525 Budapest, Hungary
5. University of Cambridge, Cavendish Laboratory, Cambridge, United Kingdom

We investigate superconducting proximity effects in NbN/Pt/Py and NbN/Cu/Py trilayers using broadband ferromagnetic resonance (FMR). The reduction of the spin-pumping contribution to the Py damping below the superconducting transition temperature (Tc) is used to probe the superconducting correlations and electronic density of state in the buried metallic interlayer. For a Cu interlayer, the damping reduction remains finite up to at least 20 nm, in agreement with the large coherence length above 100nm in Cu. In contrast, the damping reduction for a Pt interlayer exhibits a non-monotonic thickness dependence and is fully suppressed above 5 nm. Using the quasiclassical theory, we are able to explain the oscillation in damping reduction below Tc with the coexistence of superconducting and magnetic proximity effects in Pt, which generates oscillations of the superconducting order parameter and electronic density of state in the exchange-splitted Pt interlayer. Our results establish spin-pumping damping as a sensitive probe of superconducting proximity effects in buried metallic layers.

Hybrid structures combining superconductors (SC), ferromagnets (FM), and non-magnetic metals (NM) with strong spin–orbit coupling (SOC), such as Pt, have attracted considerable attention in superconducting spintronics [1-3]. Their ability to manipulate Cooper pair correlations through singlet-to-triplet conversion by the interplay between superconductivity, exchange interactions, and SOC remains debated [4-9]. And, recently, the combination of the same ingredients has emerged as a promising route toward nonreciprocal superconducting transport [10]. In particular, field-free Josephson diode behavior has been demonstrated using proximity-magnetized Pt [11], highlighting the importance of understanding how superconductivity and magnetism coexist within this metal.

In SC/NM/FM structures, Cooper pairs penetrating the NM layer, over the coherence length $\xi_{NM}$, modify its electronic density of states through the superconducting proximity effect [12]. Simultaneously, magnetic proximity effects at the NM/FM interface can induce a magnetic moment in nominally non-magnetic metals [13-15]. The resulting exchange field should generate spatially oscillating superconducting correlations in SC/NM/FM, analogous to those observed in SC/FM systems [16]. Tunneling spectroscopy [17-18] and critical current measurements in Josephson junctions [19] were used to probe the proximity-induced modification of the NM and FM quasiparticle density of states and superconducting order parameter but they require direct access to the investigated layer surface and cannot be easily applied to buried interfaces. Inverse proximity effects in buried NM layers were instead indirectly quantified by Tc measurements in Nb/Pt/Fe [20], Nb/Au/Fe [21], Nb/Au/Co [22] and Nb/Cu/CoFe [23] structures. The overall reported trend is an increase of Tc with NM interlayer thickness from 0 to $\xi_{NM}$. Additional short-period oscillations (0.75 nm) of Tc observed in single crystal Au interlayer were associated to quantum well states [24]. The absence of FFLO-like oscillations in Cu and Au interlayers was explained with the small induced moment, two orders of magnitude smaller than in FM [13, 24-26]. On the other hand, Pt is much more susceptible to magnetic polarization because it lies close to the paramagnetic instability. Consequently, hybridization with an adjacent FM layer can induce a larger magnetic moment in Pt than in Cu or Au [15,27].

Here we establish a new method based on ferromagnetic resonance (FMR) and spin pumping measurements to probe the density of states (DOS) at the Fermi level of a buried interlayers such as Pt and Cu in NbN/Pt/Py and NbN/Cu/Py stacks respectively. Especially, we reveal that the reduction of the Permalloy damping below Tc and its oscillatory behavior as a function of Pt thickness is consistent with damped oscillatory superconducting correlations induced by magnetic proximity effect. Quasiclassical calculations reproduce the main features of this thickness dependence, confirming singlet-to-triplet conversion in NbN/Pt/Py.

NbN-based heterostructures were grown by sputter deposition (PVD) on MgO(100) single crystal substrates. Our studies mainly focus on two series of samples: NbN/Pt($t_{Pt}$)/Py and NbN/Cu($t_{Cu}$)/Py where $t_{Pt}$ ranges from 0 to 18 nm and $t_{Cu}$ from 0 to 50 nm respectively. For the NbN layer we chose a thickness of approximately 50 nm, about 10 times the coherence length ($\xi_{NbN} \approx 5\ nm$ [28]), to ensure that superconductivity and Tc are only weakly affected by proximity effects, while the thickness of the Py layer was fixed at 3.5 ± 0.5 nm. A single Py(6nm) layer directly on MgO was used as reference to characterize the temperature dependence of the Gilbert damping in the absence of a SC interface. All samples were covered with MgO(2nm)/Pt(2nm) to protect the stack from oxidation while blocking spin transfer from the Py to the Pt cap. From X-ray diffraction experiments, we confirmed that bcc NbN (001) grows epitaxially on MgO(100) substrate with the epitaxial relationship $[100]_{NbN}//[100]_{MgO}$. Pt shows a single (111) texture on NbN whereas Cu contains a mixture of (100) and (110) textures. No structural information about Py can be extracted from the XRD. Thickness and interface quality were controlled by x-ray reflectivity in grazing incidence. The Tc value of each multilayer was measured around $14.6 \pm 0.4\ K$, indicating that the superconducting gap energy of NbN is similar in the whole set. The Py remanent magnetization lays in the sample plane and its saturation magnetization ($M_s$) at 300 K is $550 \pm 50\ kA/m$ for all samples. The details on the x-ray measurements, $T_c$, and $M_s$ are shown in the Supplementary material.

Broadband-FMR was performed in flip-chip geometry on a coplanar waveguide (CPW) [29] connected to a NanOsc CryoFMR setup in a QD-PPMS. The FM magnetization was put out-of-equilibrium by the combination of a static magnetic field and a rf magnetic field, perpendicular to each other in the sample plane [30]. The derivative of the transmitted power with respect to the static field was measured with a lock-in by superimposing a small AC magnetic field. The FMR resonance was detected by sweeping the field at a fixed rf frequency (from 10 to 28 GHz). The linewidth $\Delta H$ and the resonance field $H_{res}$ could be extracted from the measured spectra by a Lorentzian fit [29]. The damping constant $\alpha$ was extracted by fitting the following expression [31,32]: $\Delta H = \Delta H_0 + \frac{4\pi f \alpha}{\gamma}$ with $\Delta H_0$ being the inhomogeneous linewidth and $\gamma \approx 29.5 \pm 0.2\ GHz/T$ the gyromagnetic ratio. $\alpha$ can be defined as the sum of an intrinsic contribution of the FM $\alpha_{int}$ [33,34] and a spin pumping contribution $\alpha_{sp}$ [33,35-37] related to the pumped spin current into the adjacent layers. The eddy current contributions to the damping arising from either the sample [33] or the CPW [38] are neglected since they are at least one order of magnitude lower than $\alpha_{int}$ and $\alpha_{sp}$ [33,38,39].

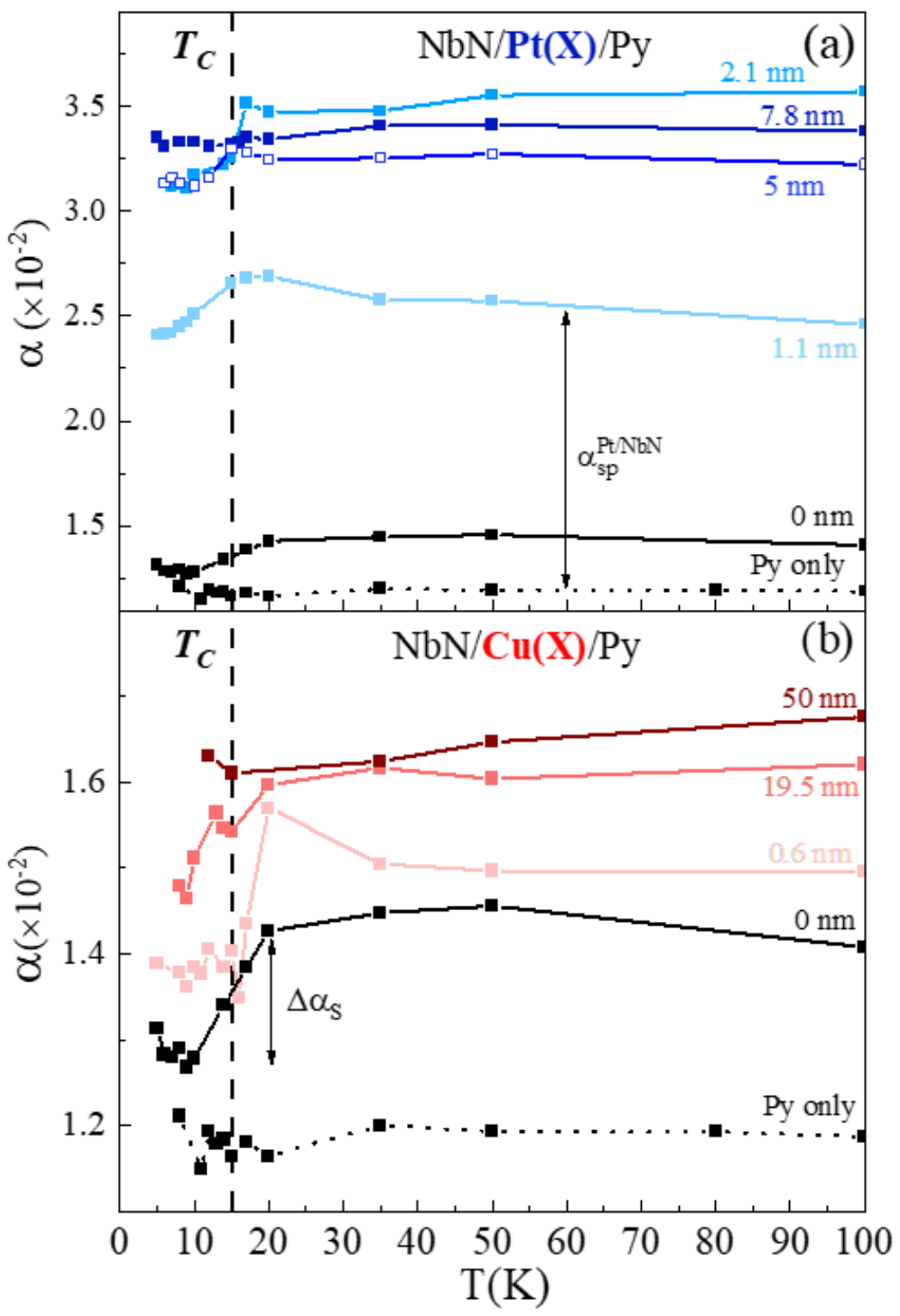

**Fig.1. Damping parameter $\alpha$ (extracted from the frequency dependence of the FMR linewidth) as a function of temperature for various interlayer thicknesses of (a) Pt and (b) Cu. The critical temperature $T_c$ is represented by a dashed line. The black arrow illustrates $\Delta\alpha^S$, the difference of damping between the normal and the superconducting states.**

By studying $\alpha_{sp}$ as a function of NbN thickness without an interlayer ($t_{Pt\ or\ Cu} = 0\ nm$) for $T > T_c$, as shown in detail in the supplementary material, we extracted the value of the effective spin-mixing conductance ($g_{\uparrow\downarrow,eff}^{Py|NbN} \approx 5.3 \pm 0.8\ nm^{-2}$) and the spin-diffusion length ($l_{sd}^{NbN} \approx 35 \pm 15\ nm$). These quantify, respectively, the efficiency of spin angular momentum transfer across the NM/FM interface and the distance over which a non-equilibrium spin polarization can propagate in the NM before it is lost due to spin-relaxation processes. The insertion of either a Pt or Cu interlayer further increases the total damping according to the nature and the thickness of the interlayer (Fig. 1). In the case of Pt, with large spin-orbit coupling, the damping increases significantly over a thickness range of a few nanometers and then saturates [40-42]. In the case of Cu with limited spin-orbit coupling, the damping increases only weakly with the thickness up to 50nm, the largest tested thickness [35,36,43]. The inset of Fig. 2 summarizes the evolution of $\alpha_{sp} (= \alpha - \alpha_{int})$ at 100K as a function of the interlayer thickness for both series of samples. When the NbN thickness is larger than $l_{sd}^{NbN}$, the NbN layer can be considered as a near-ideal spin sink. Therefore, we fitted $\alpha_{sp}$ as a function of the Pt and Cu thickness respectively at 100K using a spin pumping diffusive model for a trilayer [35,36] :

$$\alpha_{sp}(t_{NM}) = \frac{g_L \mu_B g_{\uparrow\downarrow,eff}}{4\pi M_s t_{FM}} \left( 1 + g_{\uparrow\downarrow,eff} R_{NM} \left( \frac{1 + g^* R_{NM} \tanh\left(\frac{t_{NM}}{l_{sd}^{NM}}\right)}{\tanh\left(\frac{t_{NM}}{l_{sd}^{NM}}\right) + g^* R_{NM}} \right) \right)^{-1} \quad (1)$$

where $R_{NM} = \frac{e^2}{h} l_{sd}^{NM} \rho_{NM}$ is the spin resistance of the interlayer, $\mu_B$ is the Bohr magneton, $g_L$ is the Landé factor taken at 2.1 for Py [36,44], $M_s$ is the saturation magnetization from VSM results and $t_{FM}$ the Py thickness from XRR results. $g^*$ is the effective spin transfer conductance between NM and NbN. On the one hand, we extracted $l_{sd}^{Pt} \approx 3.6 \pm 0.6\, nm$ and $g_{\uparrow\downarrow,eff}^{Py/Pt/NbN} \approx 40 \pm 5\, nm^{-2}$, consistently with values reported in the literature [45-48]. On the other hand, in the case of Cu interlayer, we extracted $l_{sd}^{Cu} \approx 112 \pm 50\, nm > l_{sd}^{NbN}$. The mixture of textures in Cu leads to a value smaller than the largest reported ones at 100K (around 1 μm) [Kim2008], and also leads to a reduced interfacial spin transparency with $g_{\uparrow\downarrow,eff}^{Py/Cu/NbN} \approx 7.8 \pm 1\, \mathrm{nm}^{-2}$ lower than 15 $nm^{-2}$ reported previously [35,36,49].

Fig. 1 shows that below Tc, the damping constant for the NbN/Py bilayer decreases concomitantly with the opening of a superconducting gap in NbN [9,39,43]. This is because the electrons participating to spin pumping have energies of the order of μeV while the NbN superconducting gap is about 2.4 meV [50,51]. Moreover, Andreev reflections producing singlet Cooper pairs do not contribute to a reduction in the damping value due to the conservation of angular momentum [43]. Note that the amplitude of the reduction $\Delta\alpha_S^{NbN}$, defined as the difference between the damping value above and below Tc, is only 85% of $\alpha_{sp}^{NbN}$. The remaining spin pumping may originate from NbN/Py interface scattering, a weakening of the superconducting order parameter near the NbN/Py interface by inverse proximity effect allowing spin dissipation [43,52] or from sample-to-sample $\alpha_{int}$ variations of Py. No significant and systematic change of $\Delta H_0$ nor $\mu_0 M_{eff}$ occurs when crossing Tc (see the Supplemental Material). In Fig. 1(a), one can observe how $\Delta\alpha_S^{Pt}$ (the loss of spin-pumping damping below Tc) varies with Pt interlayer thickness in NbN/Pt/Py. In Fig. 2, we plot $\Delta\alpha_S^{Cu}/\alpha_{sp}^{Cu}$ as a function of Cu interlayer thickness. After a first drop by 45% within the first 0.6 nm, $\Delta\alpha_S^{Cu}$ remains constant while $\alpha_{sp}^{Cu}$ monotonically increases (see inset Fig. 2 and in the Supplementary Materials). $\Delta\alpha_S^{Cu}$ remains non-null up to at least 50 nm. On the contrary, $\Delta\alpha_S^{Pt}$has a non-monotonic behavior with a local maximum at around 3.7nm before decreasing to zero for a thickness of Pt above 10 nm. The observed peak of $\Delta\alpha_S^{Pt}$ (see the Supplementary Material) together with the monotonic increase and saturation of $\alpha_{sp}^{Pt}$ leads to an exponentially damped oscillatory dependence for $\Delta\alpha_S^{Pt}/\alpha_{sp}^{Pt}$ (Fig. 2) Note that, in inset of Fig. 2, three $\alpha_{sp}^{Pt}$ values present a poor match with the fit of the spin pumping diffusive model. We have confirmed that the damped oscillation persists even when replacing the experimental values by the fit values in the calculation of $\Delta\alpha_S^{Pt}/\alpha_{sp}^{Pt}$ (see the Supplementary Material).

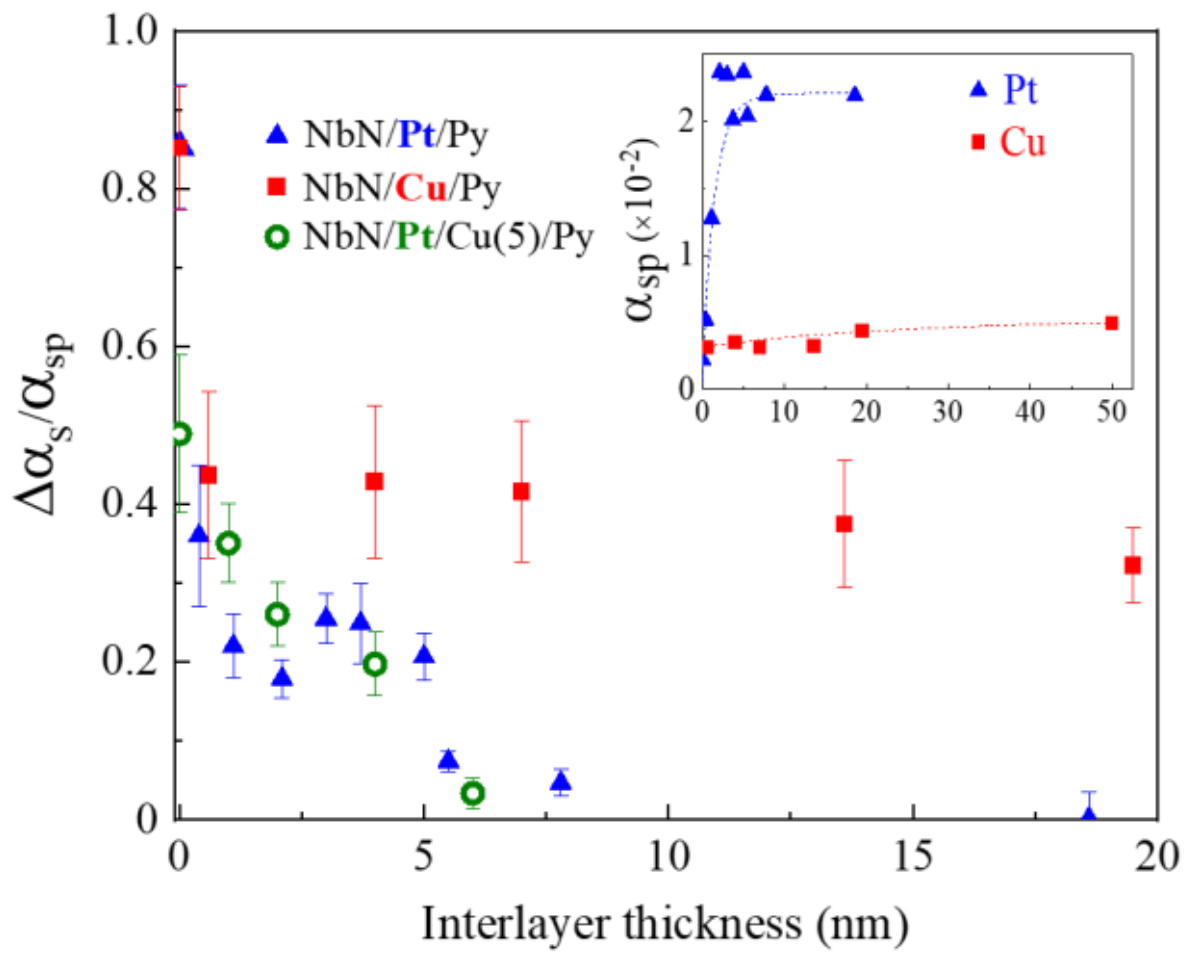

**Fig. 2 Amplitude of reduction of damping below Tc $\Delta\alpha_S$ relative to spin pumping amplitude $\alpha_{sp}$ as a function the Pt or Cu interlayer thickness in NbN/Cu/Py (red square), NbN/Pt/Py (blue triangle) and NbN/Pt/Cu(5)/Py (green circle). Inset : Spin-pumping contribution $\alpha_{sp}$ calculated as $\alpha - \alpha_{int}$ at 100K (i.e. for T > Tc) as a function of Pt thickness (blue triangles) and Cu thickness (red squares) in between NbN and Py. The dashed lines correspond to fits of the spin pumping diffusive model (Eq. (1)) to the experimental data points. The light blue and light red symbols were not included in the fits.**

A natural framework to interpret the results in Fig. 2 for the NbN/Pt/Py samples is provided by the theory of superconducting proximity effects in the presence of an exchange field where a finite center-of-mass momentum for the Cooper pairs penetrating the ferromagnetic layer results in oscillations of the local density of states and superconducting order parameter [16]. Although Pt is intrinsically non-magnetic, a moment of about 0.22 μB/at is induced at the Pt/Py interface and decays down to zero after 2 nm in Pt [15,27]. Consequently, in NbN/Pt/Py trilayers with Pt thickness below 2 nm, the Pt layer needs to be treated as a weak ferromagnet and we expect a similar phenomenology as that observed in SC/FM systems with weak ferromagnets like PdNi [19,53]. Within this picture, the oscillation of $\Delta\alpha_S^{Pt}/\alpha_{sp}^{Pt}$ with Pt thickness can be explained by the modulation of the electronic density of states at the Fermi level, which is relevant for the spin pumping from Py. An increase of Cooper pair density implies smaller damping below Tc and so a larger $\Delta\alpha_S^{Pt}$. The role of the induced moment in Pt was confirmed by growing an additional series of sample where a 5 nm-thick Cu layer was inserted between the Pt and Py layers to suppress exchange-induced effects at the interface (see the Supplemental Material). The evolution of $\Delta\alpha_S^{Pt}/\alpha_{sp}^{Pt}$ as a function of Pt thickness is plotted in Fig.2 and does not exhibit the oscillatory behavior observed when Pt and Py are in direct contact. Instead, it shows a linear monotonic decay. This is consistent with the conclusion that the Pt magnetization induced by magnetic proximity effect at the Pt/Py interface is at the origin of the oscillatory component in $\Delta\alpha_S^{Pt}/\alpha_{sp}^{Pt}$.

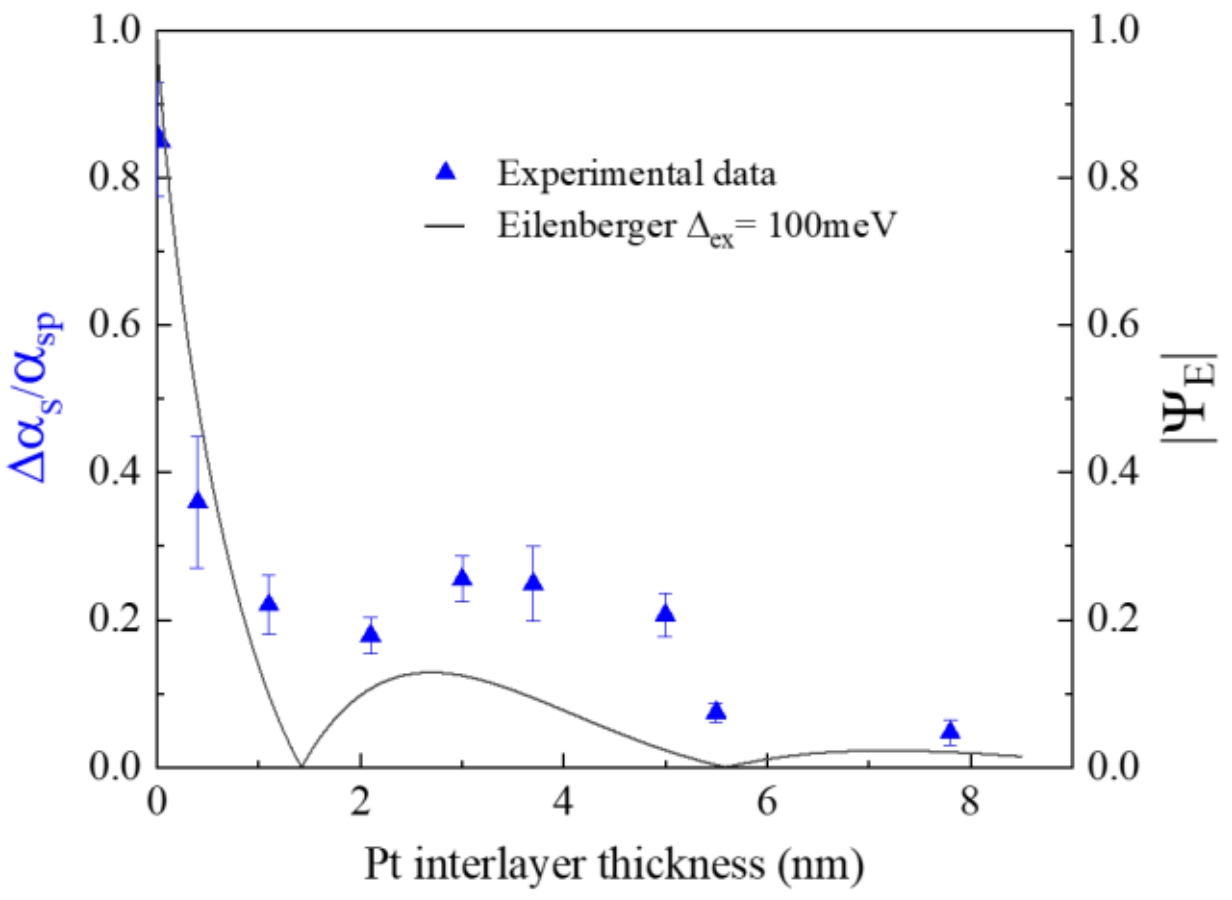

**Fig. 3. Amplitude of reduction of damping below Tc $\Delta\alpha_S$ relative to spin pumping amplitude $\alpha_{sp}$ as a function the interlayer thickness in NbN/Pt/Py (blue points), compared to the oscillation of the order parameter as a function of the Pt thickness, calculated by solving numerically the Eilenberger equation (black line).**

Oscillatory superconducting proximity effects in ferromagnets are conventionally described using the diffusive Usadel equation [54]. In this limit, the induced pair correlation has a damped-oscillatory form $\Psi(x) \propto \exp(-x/\xi_F)\cos(x/\xi_F)$ [16], where $\xi_F = \sqrt{\frac{2\hbar D}{\Delta_{ex}}}$ describes the decay and oscillation lengths. This simple expression illustrates how an exchange field causes superconducting correlations to decay in an oscillatory fashion and can therefore account qualitatively for a nonmonotonic thickness dependence. However, the relevant exchange-splitted Pt thickness must be only approximately 2 nm, substantially smaller than the estimated mean free path of $l_{mfp}$= 5 nm [55], while the largest investigated thicknesses are comparable to $l_{mfp}$. In addition, the induced spin splitting is sufficiently strong that complete momentum isotropization before appreciable exchange-induced phase accumulation cannot be assumed. We therefore use the linearized Eilenberger equation as a more accurate description of the thickness dependence. Unlike the Usadel approximation, this treatment retains the angular dependence of the quasiparticle trajectories and remains applicable across the crossover between quasiballistic and diffusive transport.

The Eilenberger equation is solved numerically using a half-space model for an effective exchange-split Pt region. The superconducting correlations are injected at the NbN/Pt boundary, while the Pt/Py interface is assumed to be transmitting and strongly dephasing, so that coherent return from Py is neglected. Elastic impurity scattering is included self-consistently through the complete angular average of the anomalous Green function (see Supplemental Information). Similarly to the oscillation of Tc in SC/FM with the FM thickness [16], we use the superconducting singlet amplitude as a representative physical quantity, correlated here to the loss of available Pt electronic density of state for FMR spin-pumping damping below Tc. This choice is further motivated by the strong spin-orbit impurity scattering in Pt, which preferentially relaxes triplet correlations while leaving conventional isotropic $s$-wave singlet correlations without a direct spin-orbit relaxation term [56]. The result for the superconducting singlet amplitude is shown in Fig. 3 for a spin-splitting $\Delta_{ex} = 100$ meV and demonstrates good agreement with the experimental measurements, capturing the experimentally observed rapid suppression and first oscillation. The latest $\Delta_{ex}$ value is consistent with both the experimental values of Pt moment at Pt/Py interface, 0.22 $\mu_B$/at [15] and of Pt exchange-field at Pt/Co interface, 170 meV [57]. It is important to note here that the comparison between the experimental data and the theoretical curve is only meaningful for Pt thickness lower than 3 nm, where most of Pt is exchange-splitted. The model cannot account for the suppression of $\Delta\alpha_S^{Pt}/\alpha_{sp}^{Pt}$ for Pt thickness larger than 5 nm ($> l_{sd}^{Pt}$) which is due to the loss of spin information inside the Pt so that the NbN superconductive transition has no impact on the FMR damping.

In conclusion, using broadband FMR, we investigated the influence of superconductivity on spin pumping in NbN/Pt/Py and NbN/Cu/Py trilayers. The superconducting transition of NbN reduces the magnetic damping by $\Delta\alpha_S^{Cu}$ and $\Delta\alpha_S^{Pt}$ respectively, revealing the transmission of superconducting correlations through the metallic interlayer, and so a reduction of available DOS at the Fermi level for spin-pumping. As the Cu interlayer thickness increases in NbN/Cu/Py, the relative damping reduction $\Delta\alpha_S^{Cu}/\alpha_{sp}^{Cu}$ first drops to 40%, possibly due to interfacial scattering, and then only exhibits a weakly thickness-dependent behavior consistent with the long coherence length of Cu. On the contrary, in NbN/Pt/Py, the initial drop is part of a damped oscillatory behavior which is suppressed when Cu is intercalated between the Pt and Py layers. This behavior is consistent with quasiclassical models of exchange-induced oscillatory superconducting correlations and associated spectral modifications in Pt. More broadly, our results establish spin-pumping spectroscopy as a sensitive probe of superconducting correlations in buried metallic layers and provide microscopic insights relevant to superconducting spintronics and the development of nonreciprocal superconducting devices.

**Acknowledgements**

This work was supported partly by the french France 2030 program «Lorraine Initiative of Excellence », reference ANR-15-IDEX-04-LUE and by France 2030 government investment plan managed by the French National Research Agency under grant references PEPR SPIN – SUPERSPIN ANR-22-EXSP-0012 and SPINCHARAC ANR-22-EXSP-008. The authors thank M. Hehn and C. Rojas-Sanchez for help with PVD and FMR set ups respectively, CCMagCryOptics@IJL for help with the magnetic and FMR measurements, CC X-gamma@IJL for help with the X-ray characterization.

**Data Availability**

All the data can be made available from the corresponding authors upon request.

**Refences**

## Supplementary Materials of "Oscillatory superconducting proximity effects in a buried Pt interlayer detected by FMR"

L. Peridon[1], T. Courtois[1], S.A Hussain[1], J.B. Tjernshaugen[2], J. Linder[2], S. ElMeligy[3], B. Újfalussy[4], K. Dumesnil[1], S. Mangin[1], S. Petit-Watelot[1], C. Ciccarelli[2,5], T. Hauet[1]

1. Université de Lorraine, CNRS, IJL, F-54000, France
2. Center for Quantum Spintronics, Department of Physics, Norwegian University of Science and Technology, NO-7491 Trondheim, Norway
3. Department of Physics, Virginia Tech, Blacksburg, Virginia 24061, USA
4. Wigner Research Centre for Physics, Institute for Solid State Physics and Optics, H-1525 Budapest, Hungary
5. University of Cambridge, Cavendish Laboratory, Cambridge, United Kingdom

## <u>Section S1 : Structural characterization</u>

The structural characterization was performed by X-ray diffraction (XRD) with a Bruker D8 Advance diffractometer in a $\theta - 2\theta$ configuration with a Cu anode ($\lambda_{Cu,K_\alpha} = 1.5406$ Å) and shown on Fig.S1.1.a for samples with Pt interlayer and on Fig.S1.1.b for samples with Cu interlayer. These measurements allow us to identify crystalline phases and analyze the crystalline quality of the whole set of samples.

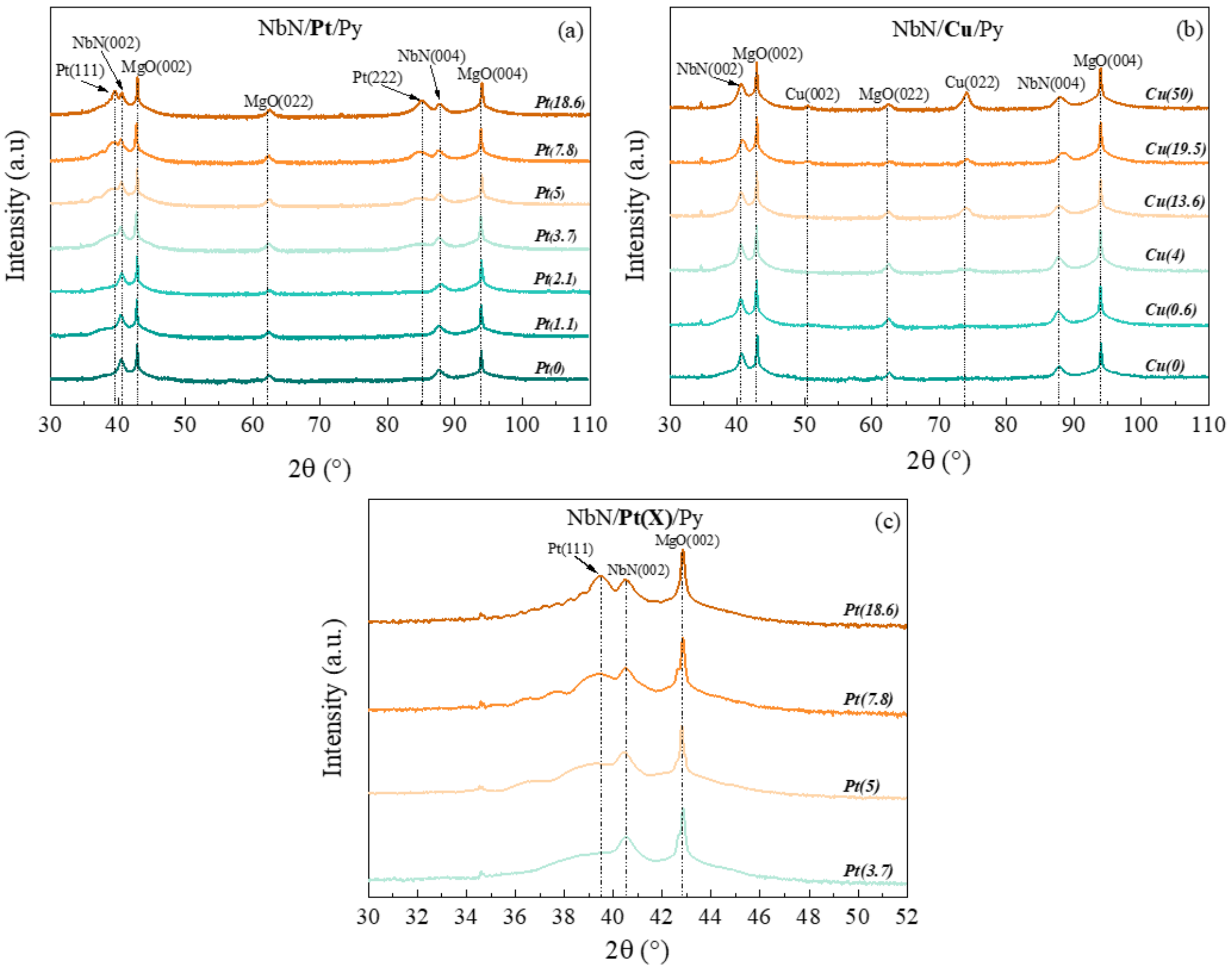

**FIG.S1.1. X-ray diffraction spectra for samples (a) MgO/MgO(5)/NbN(48-50)/Pt(X)/Py(3-3.6)/MgO(2)/Pt(2) and (b) MgO/MgO(5)/NbN(48-50)/Cu(X)/Py(3.2-3.7)/MgO(2)/Pt(2). (c) Laue oscillations visible in the diffraction spectra for samples with a Pt interlayer for four different Pt thicknesses : 3.7 / 5 / 7.8 and 18.6 nm. Dashed lines indicate the expected position of the crystalline phase.**

NbN crystallizes in the fcc structure and is well grown on a MgO(100) substrate in both types of samples. Pt has a fcc crystal structure and has a preferred (111) texture at $2\theta \approx 39.5°$ ((222) at $2\theta \approx 85.1°$) whereas Cu has a fcc crystal structure but is a mixture of two textures (100) at $2\theta \approx 50.2°$ and (110) at $2\theta \approx 73.9°$. In addition, Laue oscillations are present for the samples containing a Pt interlayer for thicknesses of 5 nm and above as shown on Fig.S1.1.c. They become increasingly pronounced as the Pt thickness increases. These oscillations evidence high structural quality of Pt on NbN. However, no structural information is obtained for the Py (111) at 43-45° or Py (001) at 51-52° due either to its polycrystallinity or a superposition with the MgO(001) peak. Consequently, it is not possible to define the Py quality on both layers of Pt and Cu.

The XRR measurements were carried out using the PANanalytical X'Pert Pro diffractometer with the samples measured at low incident angles allowing to measure the thickness of each layer. A typical curve is shown in Fig. S1.2. The tables S1.1 and S1.2 summarize the thicknesses of each layer in the NbN/Pt/Py and NN/Cu/Py samples studied in this work respectively.

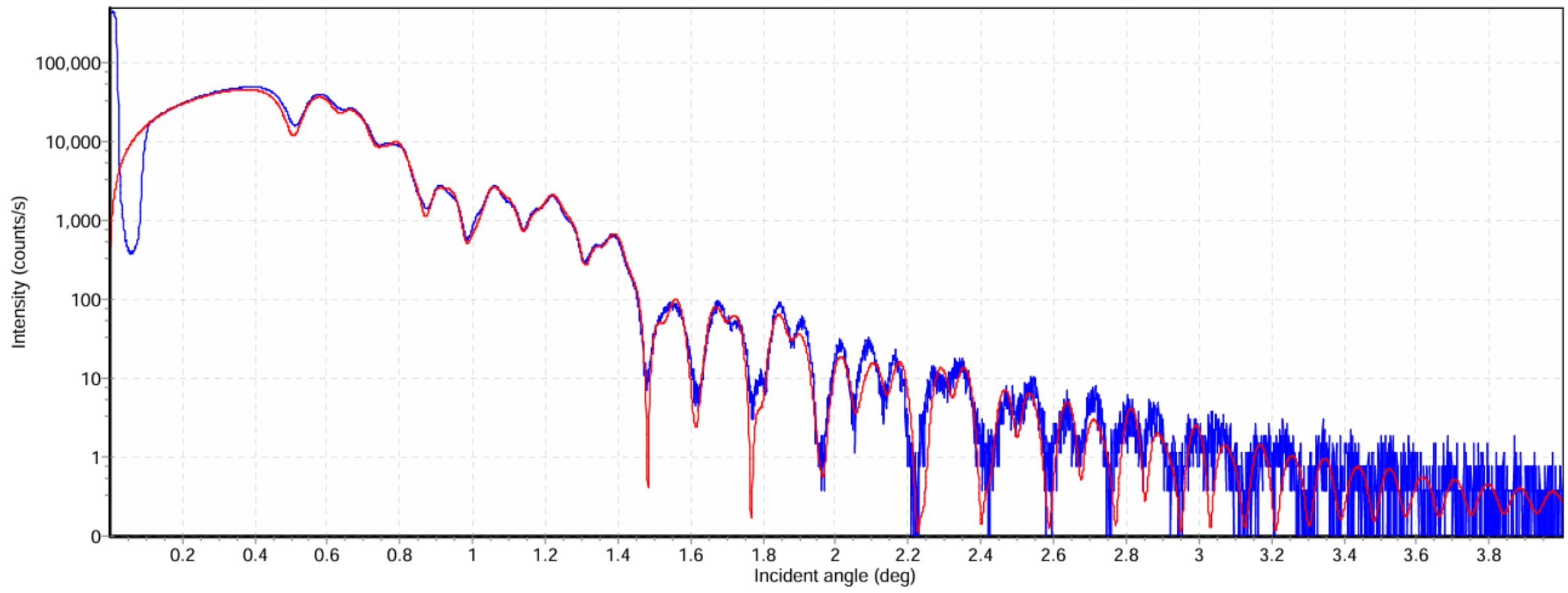

**FIG. S1.2. Reflectivity spectrum measured on the sample MgO/NbN(49)/Pt(18.6)/Py(3)/MgO(2)/Pt(2). The blue curve is the experimental curve and the red curve is the fit curve.**

| NbN thickness (nm) | Pt thickness (nm) | Py thickness (nm) |
|---|---|---|
| 49.1 | **0** | 3.3 |
| 48.5 | **0.4** | 3.4 |
| 49 | **1.1** | 3.3 |
| 49.7 | **2.1** | 3.5 |
| 47 | **3** | 4 |
| 49.8 | **3.7** | 3.6 |
| 48.9 | **7.8** | 3.4 |
| 49.1 | **18.6** | 3 |

**TABLE S1.1 : Thicknesses of the individual layers in NbN/Pt/Py samples, extracted from XRR measurements**

| NbN thickness (nm) | Cu thickness (nm) | Py thickness (nm) |
|---|---|---|
| 49.1 | **0** | 3.3 |
| 48.6 | **0.6** | 3.3 |
| 49.2 | **4** | 3.4 |
| 48 | **13.6** | 3.5 |
| 49.1 | **19.5** | 3.3 |

**TABLE S1.2 : Thicknesses of the individual layers in NbN/Cu/Py samples, extracted from XRR measurements.**

## Section S2 : Critical temperature measurements

The critical temperature $T_c$ of the samples was measured by two methods : either magnetic moments as a function of temperature $m(T)$ or resistance as a function of temperature $R(T)$. Fig. S2.1. shows the critical temperature measured for the two types of samples either with Pt or Cu. The critical temperature does not change from sample to sample and is estimated at $14.6 \pm 0.4\ K$ and thus the superconducting gap energy of NbN is the similar in the whole set of samples.

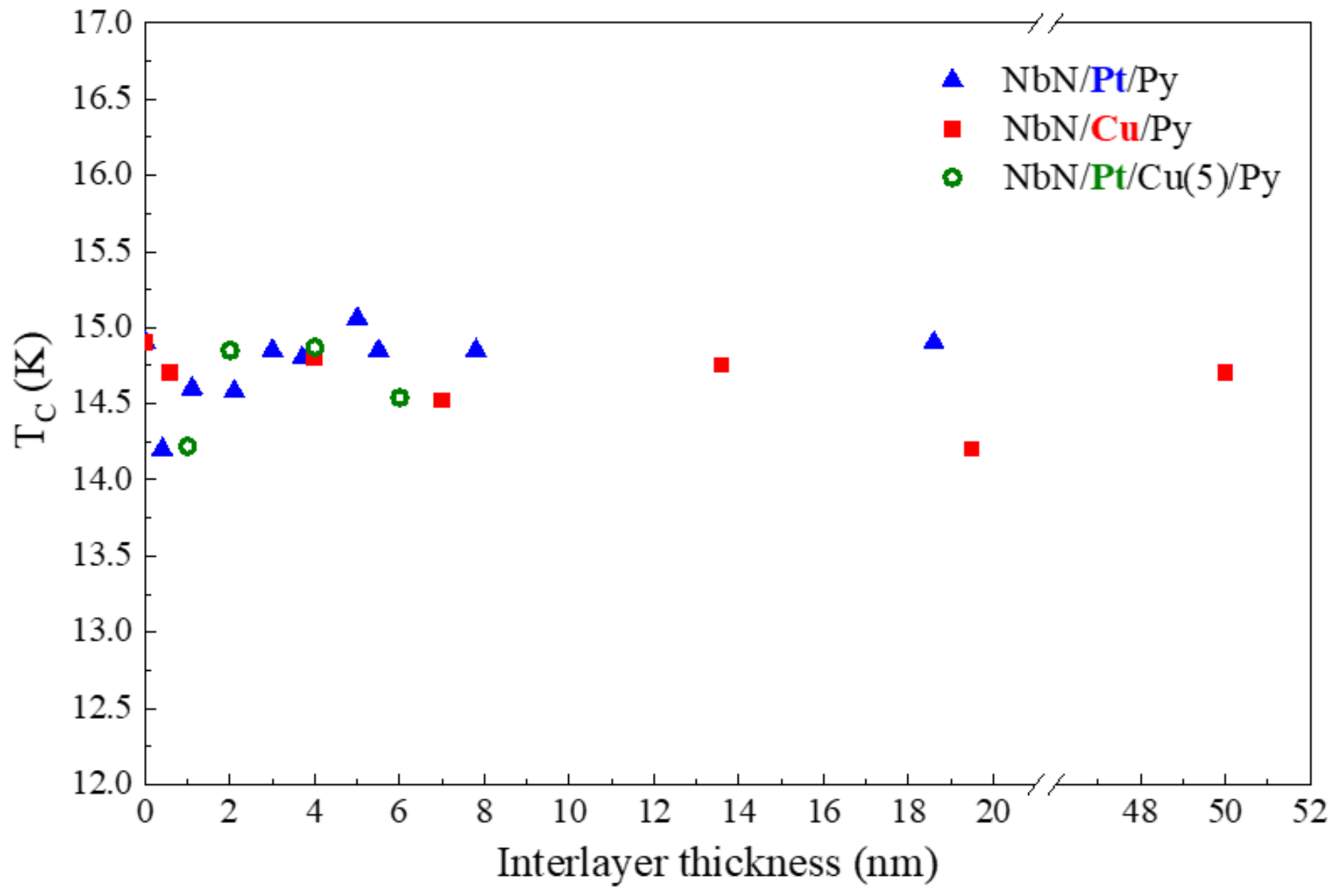

**FIG S2.1. Critical temperature $T_c$ as a function the Pt or Cu interlayer thickness in NbN/Cu/Py (red square), NbN/Pt/Py (blue triangle) and NbN/Pt/Cu(5)/Py (green circle).**

*m(T) measurements method:*

m(T) measurements were performed with a QD-Magnetometer MPMS3 SQUID (Superconducting Quantum Interference Device). The protocol measurement is as follows :

1) **Degaussing (field reset) :** Before each series of low field measurements, the residual field of the superconducting magnet is minimized using Oscillate mode (oscillate-to-zero). The field is increased to 1T and then reduced to 0T in oscillation mode. It is subsequently increased to 2 mT and again reduced to 0T in oscillation mode. This procedure minimizes the remanent field (trapped flux) and brings the system as close as possible to $H = 0T$.
2) **Zero Field Cooling (ZFC) :** At $H = 0T$, the temperature is lowered to $T_i$ well below the critical temperature of the superconductor (for instance $T_i = 3K$ for Fig. S2.2).
3) **Application of the measurement field :** In the out-of-plane (OOP) geometry where the magnetic field is applied perpendicular to the plane of the thin film, demagnetization effects are particularly significant : the internal demagnetizing field can exceed the applied field by several orders of magnitude. Therefore, a low magnetic field $H_\perp = 0.2\ mT$ is applied in order to remain predominantly in the Meissner state.
4) **ZFC Warming (red curve on Fig. S2.2) :** The m(T) curve is measured upon warming from $T_i$ to $T_f > T_c$.
5) **FC Cooling (blue curve on Fig. S2.2) :** Without changing the applied field, m(T) is then measured upon cooling from $T_f$ to $T_i$.

The ZFC Warming and FC Cooling measurements were carried out in sweep mode. The temperature was varied at a constant rate (0.5 K/min) while the magnetic moment is measured continuously throughout the temperature ramp. The critical temperature $T_c$ is determined graphically as the temperature at which the diamagnetic response disappears on the ZFC Warming curve. The associated uncertainty, indicated by the black dashed lines in Fig. S2.c., is dominated by the graphical determination, the temperature resolution, and the width of the transition; it is estimated to be $\pm 0.2K$.

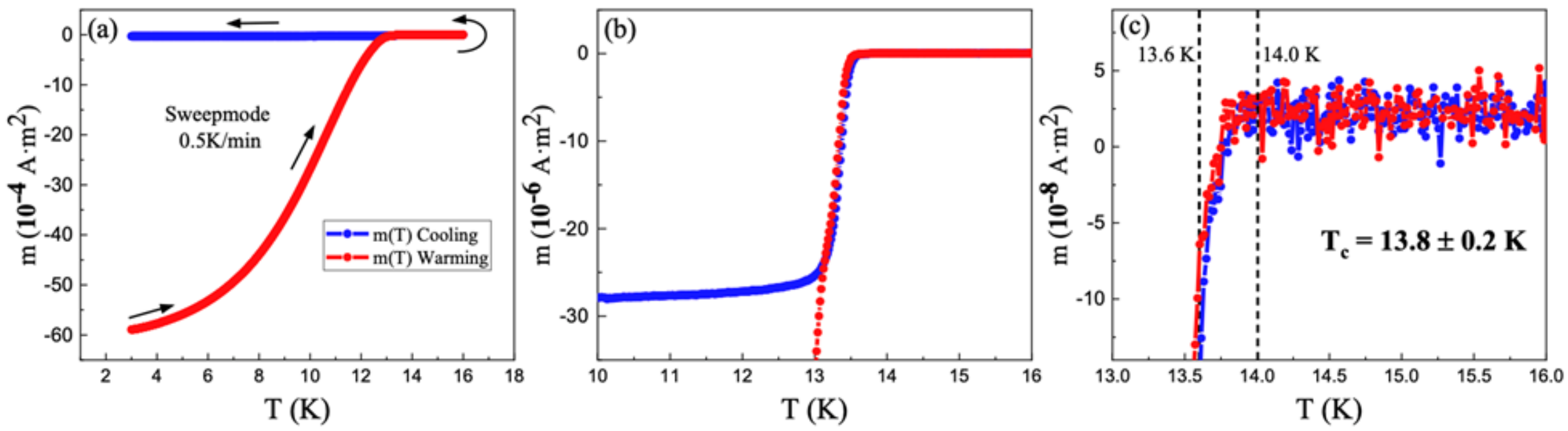

**FIG. S2.2. Determination of the critical temperature $T_c$ from the m(T) curves. (a) Magnetic moment m(T) measured under a constant low magnetic field applied perpendicular to the film ($H_\perp = 0.2\ mT$) in sweep mode (0.5 K/min), during cooling (blue curve) and warming (red curve). (b) Zoom-in on the superconducting transition region. (c) Definition of $T_c$.**

*R(T) measurements method:*

Fig. S2.3 shows a resistivity measurement performed on a MgO/MgO(5)/NbN(20)/Pt(2) sample grown using exactly the same NbN deposition recipe using a 10 $\mu$m-wide bridge. The resistive transition is sharp and does not show any particular anomaly. The $T_c$ value extracted at the 50% of the superconducting transition from this measurement is approximately 13.8 K.

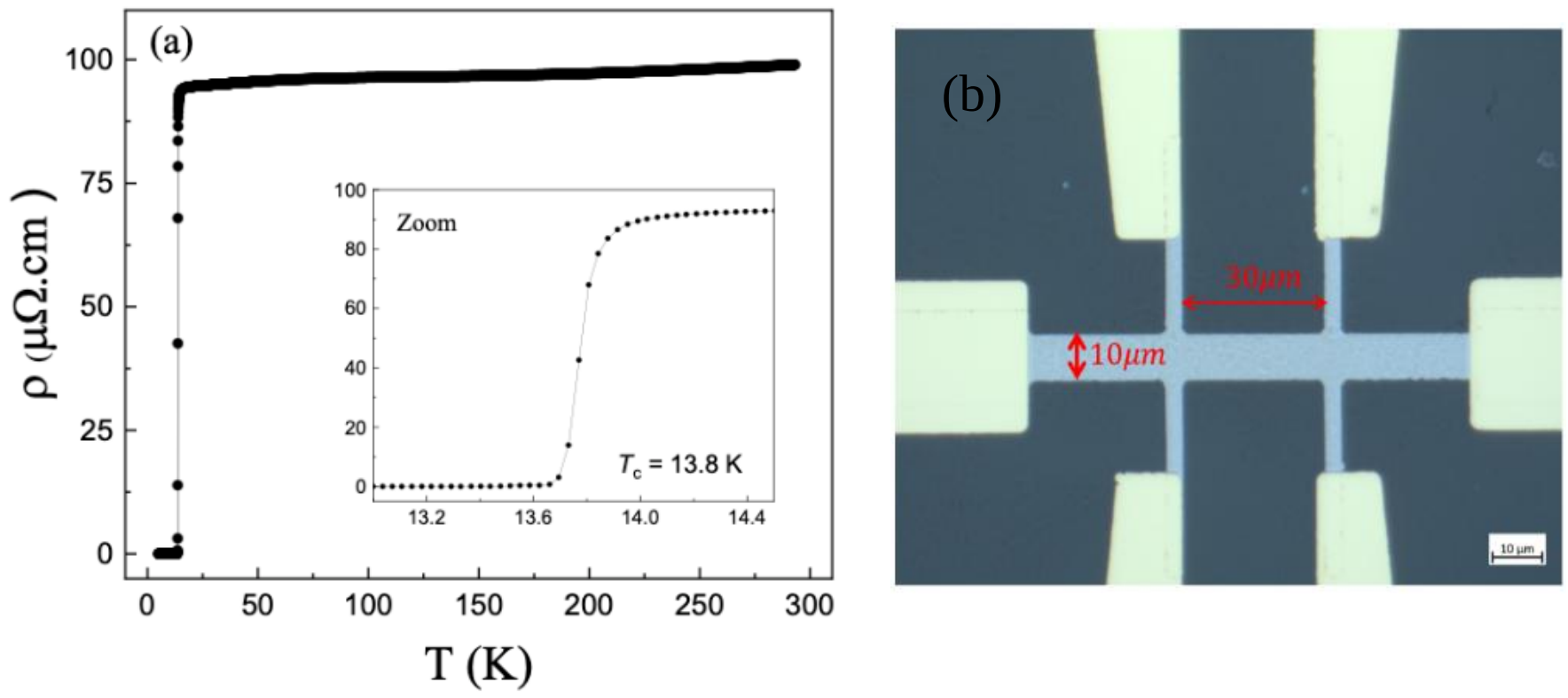

**FIG S2.3. Transport characterization of a** NbN/MgO **film ($t = 20$ nm). Measurements were performed on a microbridge ($w = 10\ \mu m$) patterned from a** NbN/MgO **film ($t = 20$ nm). (a) Resistivity $\rho(T)$ measured at zero field; the inset zooms in on the superconducting transition ($T_c \simeq 13.8$K). (b) Illustration of the microbridge.**

## Section S3 : Magnetic characterization

The magnetic characterization of the samples was performed with a MicroSense Vibrating Sample Magnetometer (VSM) at room temperature by applying an in-plane magnetic field. The fig.S3.a shows the Py saturation magnetization $M_S$ as the function of the interlayer thickness for the two types of samples. The measured Py saturation magnetization remains constant for all samples and is estimated at $550 \pm 50\ kA/m$. This value is smaller than those reported about 700 kA/m for 5-6 nm of Py in the literature [S1,S2]. All samples have a coercive field of the order of few mT as shown on fig. S3.b which represents two hysteresis loops for the two types of interlayers at equivalent thickness.

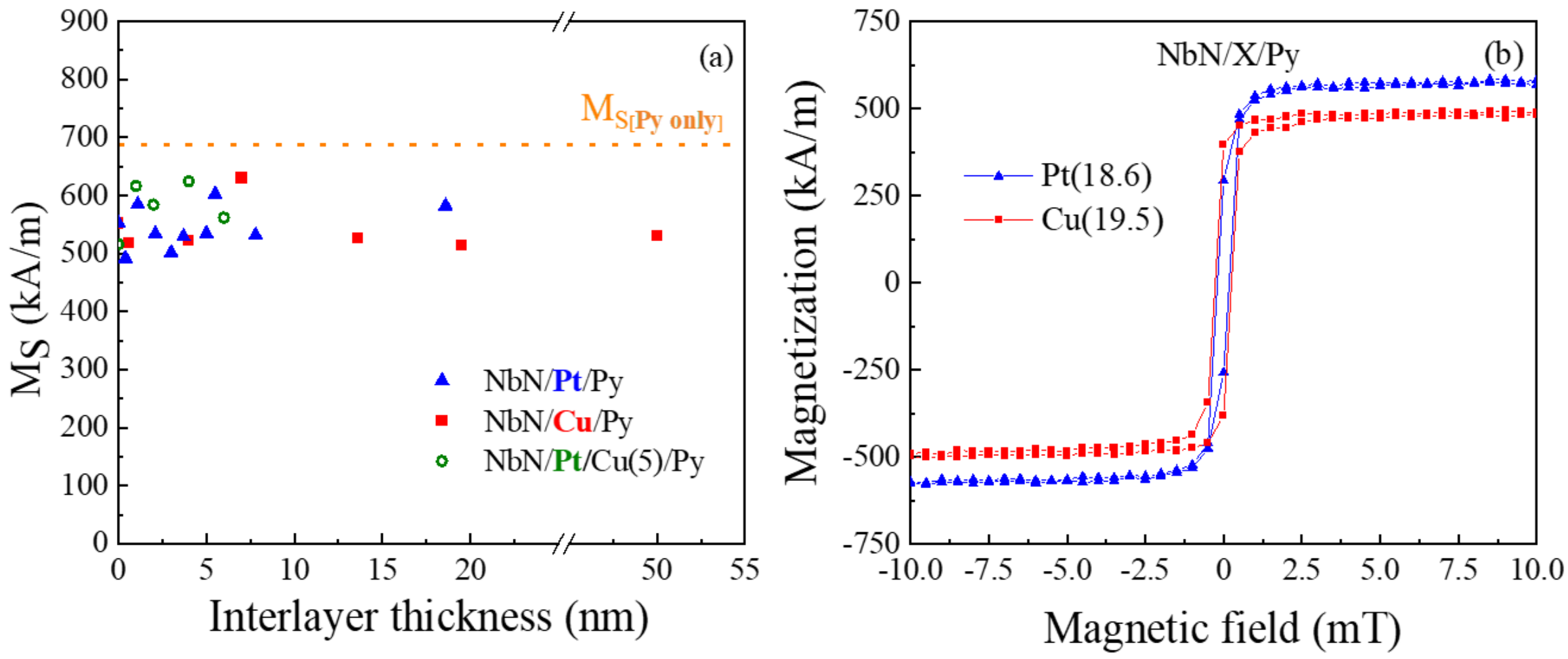

**FIG.S3. (a) Py saturation magnetization magnetization $M_S$ as a function the Pt or Cu interlayer thickness in NbN/Cu/Py (red square), NbN/Pt/Py (blue triangle) and NbN/Pt/Cu(5)/Py (green circle). The orange dashed line corresponds to the saturation magnetization of Py(6) only. (b) Hysteresis loops for an in-plane magnetic field for two different samples at equivalent interlayer thickness : Pt (blue triangles) and Cu (red squares).**

## Section S4 : $l_{sd}^{NbN}$ and $g_{\uparrow\downarrow,eff}^{Py|NbN}$ determination by FMR measurement at 100 K

To determinate $l_{sd}^{NbN}$ and $g_{\uparrow\downarrow,eff}^{Py|NbN}$ at 100K by FMR, we elaborated a series of NbN(X)/Py(3.5) samples with different NbN thicknesses : 5/10/15/30/70 nm. The presence and crystallinity of the NbN layer on MgO(100) substrate were confirmed by XRD under the same experimental conditions as those described in Section S1 and shown in Fig.S4.a. We fitted $\alpha_{sp}^{NbN}(= \alpha - \alpha_{int})$ at 100 K as a function of the NbN thickness using a spin pumping model for a bilayer [S3,S4]: $\alpha_{sp}(t_N) = \frac{g_L \mu_B g_{\uparrow\downarrow,eff}}{4\pi M_s t_{FM}}\left(1 + \frac{g_{\uparrow\downarrow,eff} R_N}{\tanh\left(\frac{t_N}{l_{sd}^N}\right)}\right)^{-1}$ where $g_{\uparrow\downarrow,eff}$ is the effective spin-mixing conductance, $R_N$ is the spin resistance of the interlayer, $\mu_B$ is the Bohr magneton, $g_L$ is the Landé factor taken at 2.1 for Py [S4,S5], $M_s$ is measured at $450 \pm 25\ kA/m$ by VSM at 300K and $t_{FM}$ the ferromagnet thickness is taken from XRR results. We extracted $l_{sd}^{NbN} \approx 35 \pm 15\ nm$ and $g_{\uparrow\downarrow,eff}^{Py|NbN} \approx 5.3 \pm 0.8\ nm^{-2}$. Our extracted spin-diffusion length of NbN at 100K from FMR is higher than that estimated at $l_{sd}^{NbN} \approx 14 \pm 3\ nm$ at 300K from spin pumping experiment with a polycrystalline NbN in a YIG/NbN bilayer [S6].

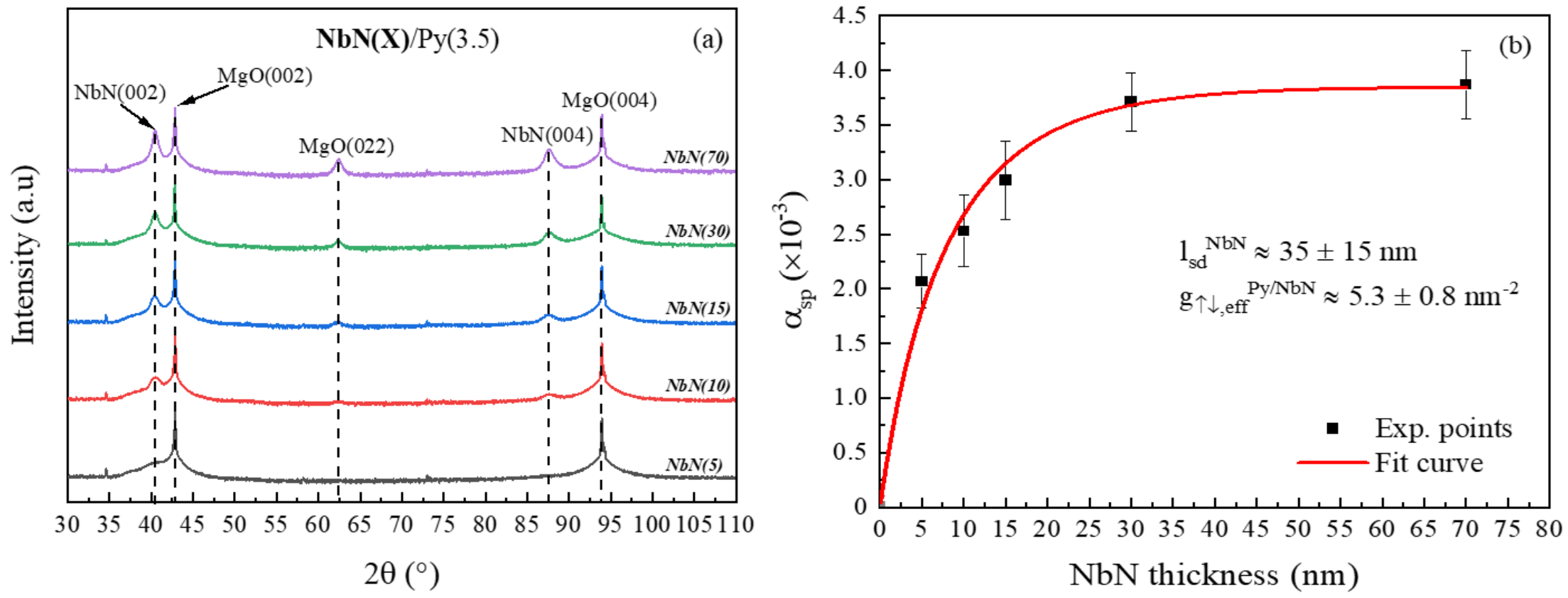

**FIG.S4. (a) X-ray diffraction spectra for samples MgO/MgO(5)/NbN(X)/Py(3.5)/MgO(2)/Pt(2). Dashed lines indicate the expected position of the crystalline phase. (b) Damping parameter $\alpha$ (extracted from the frequency dependence of the FMR linewidth) as a function of temperature (black squares) and corresponding fit of the bilayer model to the experimental data points (red curve).**

## Section S5 : $\Delta\alpha_S^{Cu}$ and $\Delta\alpha_S^{Pt}$ vs Cu and Pt interlayer thickness respectively

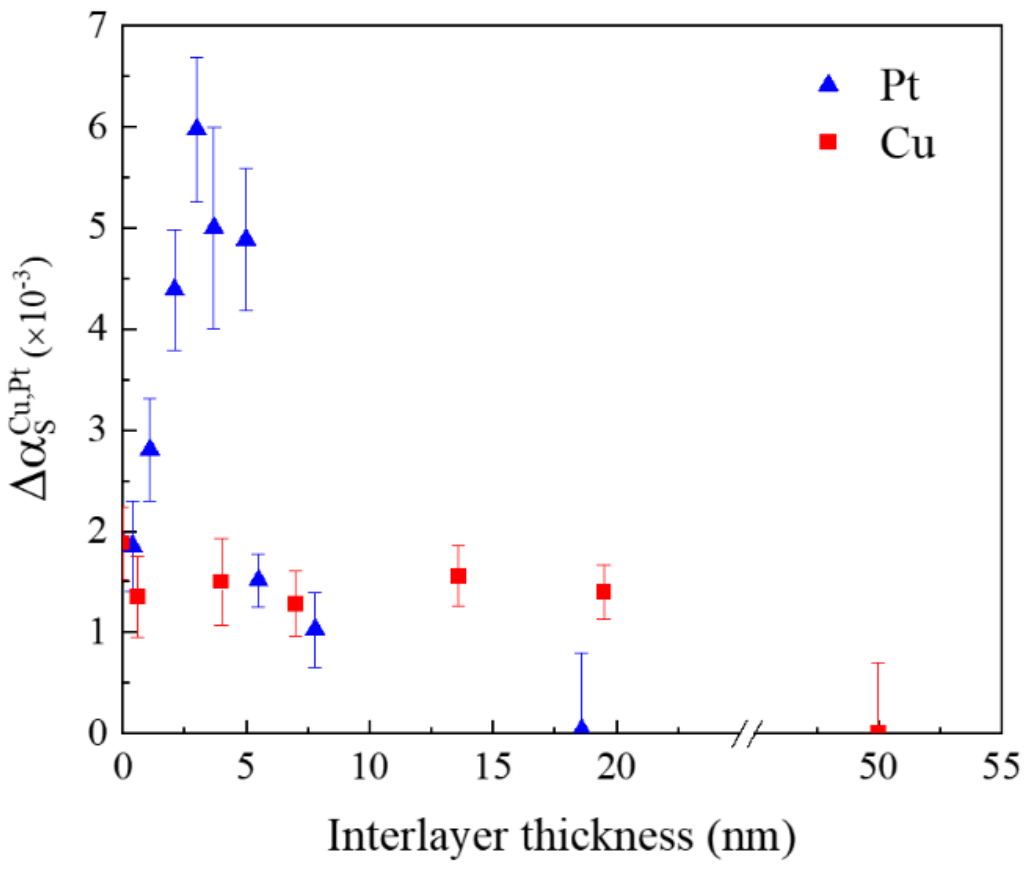

**Fig. S5. Amplitude of reduction of damping below Tc $\Delta\alpha_S$ as a function the interlayer thickness in NbN/Cu/Py (red points) and NbN/Pt/Py (blue points).**

## Section S6 : $\Delta H_0$ and $M_{eff}$ as a function of temperature, for various Cu and Pt interlayer thicknesses

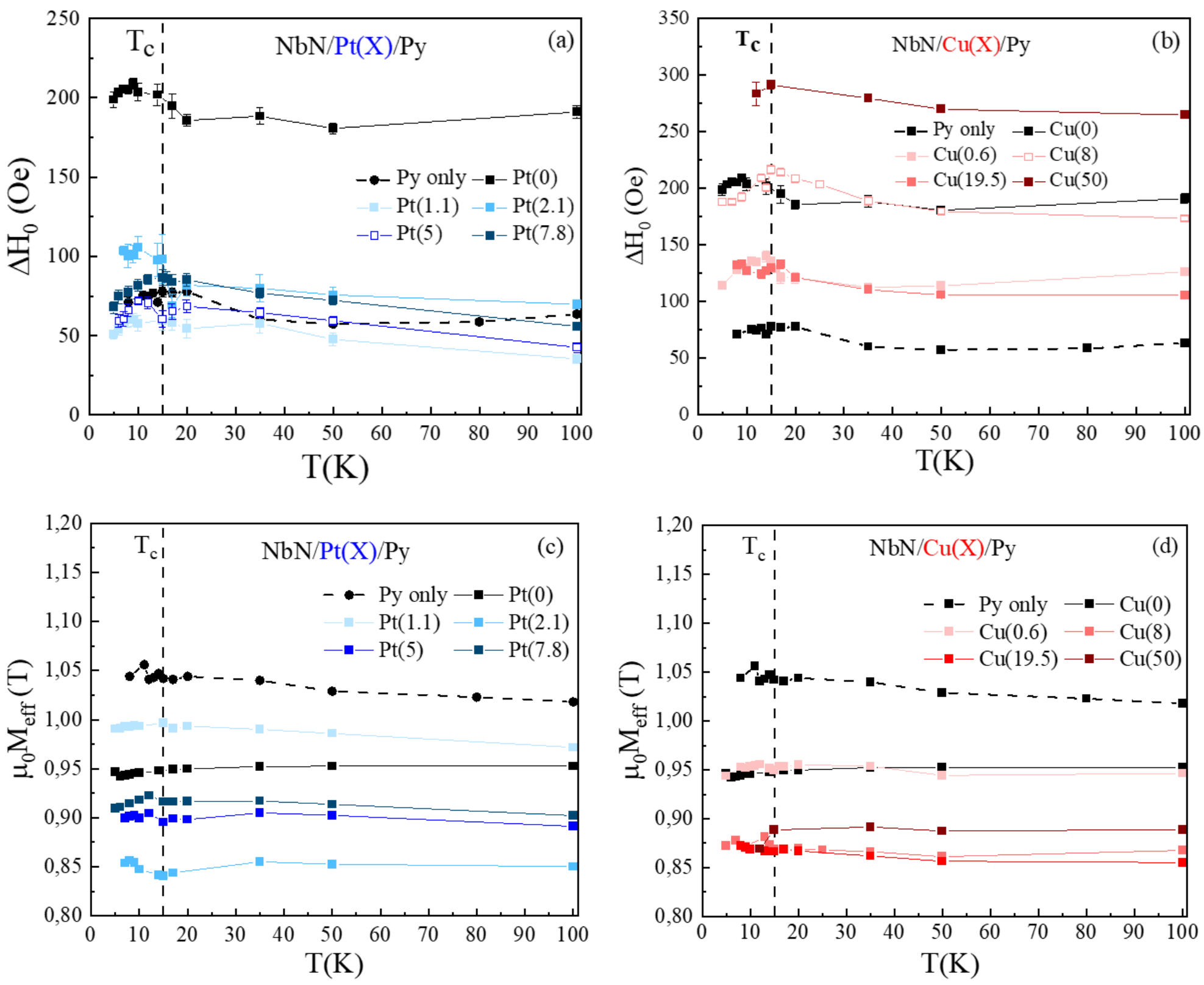

**Fig.S6. Inhomogeneous linewidth $\Delta H_0$ (extracted from the frequency dependence of the resonance linewidth) and effective magnetization $\mu_0 M_{eff}$ (extracted from the frequency dependence of the resonance field) as a function of temperature for various interlayer thicknesses of (a,c) Pt and (b,d) Cu. The critical temperature $T_c$ is represented by a dashed line. No significant and systematic change of $\Delta H_0$ nor $\mu_0 M_{eff}$ occurs when crossing Tc.**

## Section S7 : $\Delta\alpha_S^{Pt}/\ \alpha_{sp}^{Pt}$ as a function of the Pt thickness with the calculated values of $\alpha_{sp}^{Pt}$

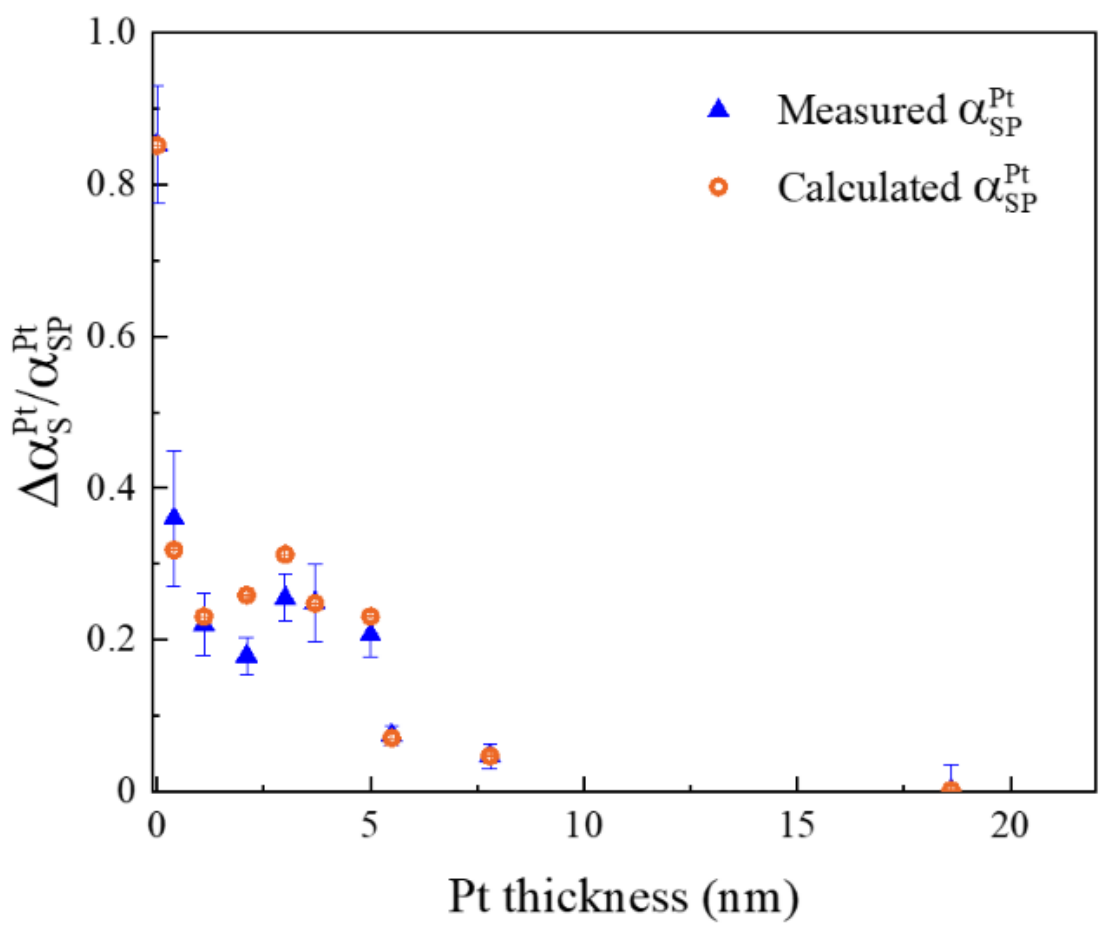

**Fig.S7. Amplitude of reduction of damping below Tc $\Delta\alpha_S^{Pt}$ relative to spin pumping amplitude $\alpha_{sp}^{Pt}$ as a function the Pt interlayer thickness in NbN/Pt/Py by using the measured $\alpha_{sp}^{Pt}$ (blue triangle) and the calculated $\alpha_{sp}^{Pt}$ extracted from the fit of the spin pumping diffusive model to the experimental points (orange circle). Both curves exhibit a damped oscillatory behavior although the minimum of the theoretical curve is slightly shifted toward a lower Pt thickness compared with the experimental curve. Therefore, that oscillation does not originate from the experimental variations of $\alpha_{sp}^{Pt}$.**

**Section S8 : $\alpha$, $\Delta H_0$ and $M_{eff}$ as a function of temperature of NbN/Pt(X)/Cu(5)/Py samples for various Pt interlayer thicknesses**

Figure S8(a) represents the damping values as a function the temperature for the NbN/Pt(X)/Cu(5)/Py for various Pt thicknesses. The insertion of a 5 nm-thick Cu layer between the Pt and Py layers considerably reduces the damping compared with NbN/Pt(X)/Py samples by suppressing the direct Py/Pt interface and reducing the effective spin-mixing conductance of the multilayer such as $7.8\ nm^{-2}\left(= g_{\uparrow\downarrow,eff}^{Py|Cu}\right) \leq g_{\uparrow\downarrow,eff}^{Py|Cu/Pt} \leq 40\ nm^{-2}\left(= g_{\uparrow\downarrow,eff}^{Py|Pt}\right)$. This suggests that more spins can diffuse in the Cu layer because $t_{Cu}(5nm) < l_{sd}^{Cu}(112nm)$ and the effect of the Pt layer known as an ideal spin-sink, is strongly attenuated. In the normal state (T > Tc), the damping remains nearly independent of the Pt thickness over the investigated range from 1 to 6 nm, suggesting that the $l_{sd}^{Pt}$ is already reached and thus is shorter than that extracted with the spin pumping diffusive model **[S3]** for NbN/Pt(X)/Py samples ($l_{sd}^{Pt} = 3.6 \pm 0.6\ nm$). In the superconducting state (T < Tc), the damping decreases as the superconducting gap opens in NbN, reducing the accessible DOS in NbN layer but also in proximized adjacent layers. For a Pt thickness larger than 6 nm, the damping in the superconducting state follows the same behavior as in the normal state meaning that the spin current is no longer affected by the superconducting proximity effect in the Pt layer.

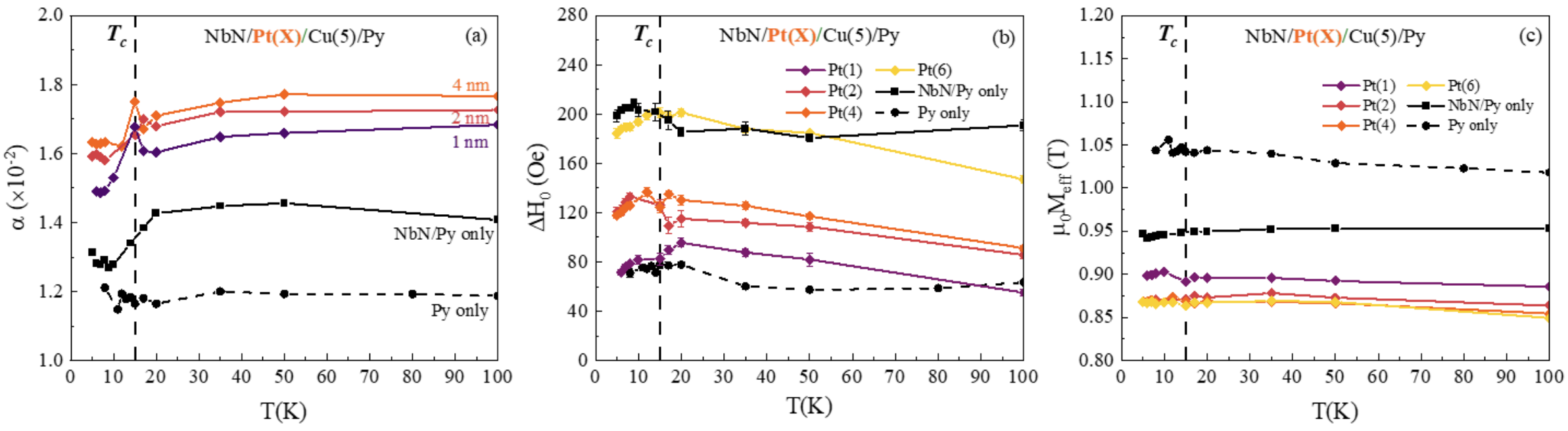

**Fig.S8. (a) Damping $\alpha$ (extracted from the frequency dependence of the resonance linewidth), (b) inhomogeneous linewidth $\Delta H_0$ (extracted from the frequency dependence of the resonance linewidth) and (c) effective magnetization $\mu_0 M_{eff}$ (extracted from the frequency dependence of the resonance field) as a function of temperature of NbN/Pt(X)/Cu(5)/Py samples for various interlayer thicknesses of Pt. The addition of 5 nm-thick Cu layer was inserted between the Pt and Py layers to suppress the spin polarization induced in Pt. The critical temperature $T_c$ is represented by a dashed line. No significant and systematic change of $\Delta H_0$ nor $\mu_0 M_{eff}$ occurs when crossing Tc.**

**Section S9 : Calculation of the order parameter in exchanged-splitted Pt with the Eilenberger equation**

To model the superconducting proximity effect in Pt/Py, we describe the superconducting correlations propagating through an effective exchange-split Pt region. The Pt/Py interface is assumed to be sufficiently transmitting and strongly dephasing that coherent return trajectories from Py can be neglected. First, we give a solution which describes one-way propagation from the superconducting interface. Then, we give a more accurate solution which includes mixing of trajectories due to impurity scattering. The latter one is used to fit the experimental data.

The linearized Eilenberger equation [S8] for the anomalous Green function $f_+(x, \mu, \omega_n) = f_s + f_t$ is

$$v_F \mu \partial_x f_+ + \frac{2(\omega_n + ih)}{\hbar} f_+ = -\frac{1}{\tau}(f_+ - \langle f_+ \rangle). \qquad (1)$$

Here, $\mu = \cos\theta$, $\omega_n = (2n+1)\pi k_B T$, $h$ is the exchange-field energy, $\tau$ is the elastic scattering time, and $l_{\mathrm{mfp}} = v_F \tau$ is the mean free path. The complementary spin sector has an equation for $f_- = f_s - f_t$, which is obtained from above simply by letting $h \rightarrow -h$. Here, $f_\pm = f_s \pm f_t$ denote the two exchange eigencombinations of the singlet and $S_z = 0$ triplet amplitudes, with the spin quantization axis chosen along the exchange field. We note in passing that $f_- = f_+^*$ holds, such that

$$f_s = \frac{f_+ + f_-}{2} = \mathrm{Re}\, f_+\,,\; f_t = \frac{f_+ - f_-}{2} = i\,\mathrm{Im}\, f_+\,. \qquad (4)$$

The notation

$$\langle f_+ \rangle = \frac{1}{2}\int_{-1}^{1} d\,\mu\, f_+$$

means average over the Fermi surface. We use the convention

$$\Delta_{\mathrm{ex}} = 2h, \qquad (2)$$

such that the exchange-induced wave vector is

$$\beta = \frac{\Delta_{\mathrm{ex}}}{\hbar v_F} = \frac{2h}{\hbar v_F}. \qquad (3)$$

The measured superconductivity-induced change in damping is such that the damping in the superconducting state is smaller than in the normal state throughout the investigated thickness range. However, the DOS in an S/F bilayer with triplet Cooper pairs will generally oscillate between exceeding the low-energy normal-state DOS and being smaller than it, depending on the distance from the interface. From the experimental measurements, we thus infer that the singlet correlations are the most important for the behavior of the damping. We therefore use the singlet component, which is obtained from the real part of $f_+$, as the simplest sign-consistent proxy for the observed thickness dependence.

Note that this does not imply that the accompanying $S_z = 0$ triplet component is generally absent. Instead, a possible reason for the singlet dominance is the known fact that spin-orbit scattering causes triplets to decay faster, while conventional isotropic $s$-wave singlet correlations are unaffected [S9]. The strong spin-orbit interaction in Pt may therefore preferentially attenuate the induced triplet component, justifying that we focus on the singlet part.

**Solution without $\langle f \rangle$ term.**

To obtain a compact analytical result, we first neglect the impurity scattering term $\langle f_+ \rangle/\tau$. We refer to this as the single trajectory solution. The term "single trajectory" refers to the absence of impurity-induced angular redistribution and return trajectories. We still keep the dephasing of the correlations due to impurity scattering. Equation (1) then reduces to

$$v_F \mu \partial_x f_+ + \left[\frac{2\omega_n}{\hbar} + i\frac{\Delta_{\mathrm{ex}}}{\hbar} + \frac{1}{\tau}\right] f_+ = 0. \qquad (5)$$

For trajectories injected from the superconducting side, $\mu > 0$, we impose

$$f_+(x = 0) = f_0(\omega_n), \qquad (6)$$

with the BCS source amplitude

$$f_0(\omega_n) = \frac{\Delta}{\sqrt{\omega_n^2 + \Delta^2}}. \qquad (7)$$

The trajectory-resolved solution of Eq. (5) is

$$f_+(x, \mu, \omega_n) = f_0(\omega_n)\exp\left[-\frac{(\alpha_n + i\beta)x}{\mu}\right], \qquad (8)$$

where

$$\alpha_n = \frac{1}{l_{\mathrm{mfp}}} + \frac{2\omega_n}{\hbar v_F},\ \beta = \frac{\Delta_{\mathrm{ex}}}{\hbar v_F}. \qquad (9)$$

Equation (8) is exact for the reduced single trajectory transport equation and does not require $x \gg \xi_h$, where

$$\xi_h = \frac{1}{\beta} = \frac{\hbar v_F}{\Delta_{\mathrm{ex}}}. \qquad (10)$$

The trajectory-resolved singlet component is the real part, hence

$$f_s(x, \mu, \omega_n) = f_0(\omega_n) \exp\left(-\frac{\alpha_n x}{\mu}\right) \cos\left(\frac{\beta x}{\mu}\right). \qquad (11)$$

We define the corresponding angularly averaged singlet correlation, which is a measure of the strength of the singlet superconducting correlations at a given position and energy, as follows:

$$F_s^{(0)}(x, \omega_n) = f_0(\omega_n) \int_0^1 d\mu\ \exp\left(-\frac{\alpha_n x}{\mu}\right) \cos\left(\frac{\beta x}{\mu}\right). \qquad (12)$$

Here, the superscript $(0)$ denotes the leading solution obtained by neglecting the impurity scattering-in term $\langle f_+ \rangle / \tau$. In this approximation, no backward-going correlations are generated, and the incoming boundary condition at the remote side gives $f_+^{(0)}(\mu < 0) = 0$. It is therefore sufficient to integrate over the forward-going trajectories, $\mu > 0$. This expression explicitly retains the different exchange-induced phases accumulated by trajectories with different incidence angles. In particular, oblique trajectories have a longer propagation path $x/\mu$ and therefore accumulate phase more rapidly than normally incident trajectories. Their destructive interference could therefore be expected to generate a first zero of the angular average earlier than the first zero of the large-distance sine asymptote at $x = \pi \xi_h$. We will see below that this is exactly the case, which improves the fit to the experimental data.

The angular integral in Eq. (12) can be expressed analytically. Introducing $t = 1/\mu$ gives

$$F_s^{(0)}(x, \omega_n) = f_0(\omega_n) \mathrm{Re} \int_1^\infty \frac{dt}{t^2}\ \exp[-(\alpha_n + i\beta) x t]. \qquad (13)$$

We define the generalized exponential integral

$$E_2(z) = \int_1^\infty \frac{dt}{t^2} e^{-zt}. \qquad (14)$$

The angularly averaged singlet correlation is therefore

$$F_s^{(0)}(x, \omega_n) = f_0(\omega_n)\, \mathrm{Re}\, E_2\, [(\alpha_n + i\beta)x]. \qquad (15)$$

Equation (15) is valid for arbitrary $x/\xi_h$ under the assumption that $\langle f \rangle$ can be neglected, which we will comment on below. In particular, it satisfies

$$F_s^{(0)}(0, \omega_n) = f_0(\omega_n), \qquad (16)$$

since $E_2(0) = 1$.

The final result for our singlet superconducting amplitude, summed over Matsubara frequencies, is then

$$\Psi_s^{(0)}(x) = \mathcal{N} \sum_{0<\omega_n<\omega_c} \frac{\Delta}{\sqrt{\omega_n^2 + \Delta^2}} \mathrm{Re}\, E_2 \left[ \left( \frac{1}{l_{\mathrm{mfp}}} + \frac{2\omega_n}{\hbar v_F} + i \frac{\Delta_{\mathrm{ex}}}{\hbar v_F} \right) x \right], \qquad (17)$$

where $\omega_c$ is the pairing cutoff and $\mathcal{N}$ is an overall normalization factor. Only a summation over positive frequencies is needed since the singlet correlator is even in frequency due to the Pauli principle.

**Numerical results.**

To assess the effect of impurity-induced angular mixing, we solve Eq. (1) numerically with the scattering term $\langle f_+ \rangle / \tau$ retained. For each positive Matsubara frequency, we discretize the trajectory direction $\mu$ and the spatial coordinate $x$. Trajectories with $\mu > 0$ are propagated forward from the superconducting boundary, where $f_+(0, \mu > 0, \omega_n) = f_0(\omega_n)$, while trajectories with $\mu < 0$ are propagated backward from a remote absorbing boundary.

Starting from the single trajectory solution, we calculate

$$\langle f_+(x, \omega_n) \rangle = \frac{1}{2} \int_{-1}^{1} d\,\mu\, f_+(x, \mu, \omega_n), \qquad (22)$$

use this angular average as the source term in the trajectory equations, and iterate until self-consistency is reached. After convergence, the local isotropic singlet correlation is obtained from the complete Fermi-surface average,

$$F_s(x, \omega_n) = \mathrm{Re} \left[ \frac{1}{2} \int_{-1}^{1} d\,\mu\, f_+(x, \mu, \omega_n) \right]. \qquad (23)$$

The result is then summed over positive Matsubara frequencies and normalized to its own value at the superconducting interface. The numerical solution is shown in Fig. 3 in the main manuscript and reproduces the rapid initial suppression and places the first minimum close to the experimental value using $\Delta_{\mathrm{ex}} = 100$ meV.